\documentclass[notitlepage,aps,pra,twocolumn,groupaddress]{revtex4-1}

\usepackage[left=2cm,top=2cm,right=2cm,headsep=0.5cm,headheight=0.5cm,nofoot]{geometry}

\usepackage{setspace,mathrsfs,mathpazo,avant,tabularx}
\usepackage[utf8]{inputenc}
\usepackage[table]{xcolor}
\usepackage{colortbl}
\usepackage{indentfirst}
\usepackage{stmaryrd}
\usepackage{amssymb,amsmath,amsthm,amsfonts,amsbsy}
\usepackage{bm,bibunits,color,chngcntr,epsfig,epstopdf,graphicx,dsfont}
\usepackage{hyperref,lipsum,,makecell,mathrsfs,rotating}
\usepackage[english]{babel}
\usepackage[normalem]{ulem}

\newcommand{\be}{\begin{equation}}
\newcommand{\ee}{\end{equation}}
\newcommand{\bea}{\begin{eqnarray}}
\newcommand{\eea}{\end{eqnarray}}
\newcommand{\ket}{\rangle}
\newcommand{\bra}{\langle}

\newcommand{\I}{\mathds{1}}
\newcommand{\ra}{\rightarrow}

\def\C#1{\mathcal #1}
\def\B#1{\mathbb #1}

\definecolor{gray}{gray}{0.9}

\begin{document}
\newtheorem{theorem}{Theorem}
\newtheorem{prop}[theorem]{Proposition}
\newtheorem{corollary}[theorem]{Corollary}
\newtheorem{open problem}[theorem]{Open Problem}
\newtheorem{conjecture}[theorem]{Conjecture}
\newtheorem{definition}{Definition}
\newtheorem{remark}{Remark}
\newtheorem{example}{Example}
\newtheorem{task}{Task}

\title{Universal resources for quantum computing}
\author{Dong-Sheng Wang}\thanks{wds@itp.ac.cn}
\affiliation{CAS Key Laboratory of Theoretical Physics, Institute of Theoretical Physics,
Chinese Academy of Sciences, Beijing 100190, China}
\date{\today}



\begin{spacing}{1.2}

\begin{abstract}
Unravelling the source of quantum computing power has been a major goal 
in the field of quantum information science.
In recent years, the quantum resource theory (QRT) has been established
to characterize various quantum resources,
yet their roles in quantum computing tasks still require investigation.
The so-called universal quantum computing model (UQCM),
e.g., the circuit model, 
has been the main framework to guide the design of quantum algorithms, 
creation of real quantum computers etc.
In this work, we combine the study of UQCM together with QRT.
We find, on one hand,
using QRT can provide a resource-theoretic characterization of a UQCM,
the relation among models and inspire new ones,
and on the other hand,
using UQCM offers a framework to apply resources,
study relation among resources and classify them.

We develop the theory of \emph{universal resources} in the setting of UQCM,
and find a rich spectrum of UQCMs and the corresponding universal resources. 
Depending on a hierarchical structure of resource theories,
we find models can be classified into \emph{families}.
In this work, we study three natural families of UQCMs in details:
the amplitude family, 
the quasi-probability family,
and the Hamiltonian family. 
They include some well known models, like 
the measurement-based model and adiabatic model, 
and also inspire new models such as the contextual model we introduce.
Each family contains at least a triplet of models, 
and such a succinct structure of families of UQCMs offers a unifying picture 
to investigate resources and design models.
It also provides a rigorous framework to resolve puzzles, such as
the role of entanglement vs. interference, 
and unravel resource-theoretic features of quantum algorithms.
\end{abstract}

\maketitle

\section{Introduction}

\subsection{Motivation}

The attempt to identify the key resource for quantum speedup
has ever been started at the birth of quantum computing. 
With notable discoveries such as the quantum teleportation~\cite{BBC+93},
quantum \emph{entanglement} was recognized to be a unique feature~\cite{JL03,Ste03}. 
At the meantime, 
from the study of early quantum algorithms~\cite{Sho94,KSV02} 
and the quantum circuit model~\cite{NC00},
it was proposed quantum \emph{interference} is the source of power for quantum computing~\cite{CEM+98}.
A quantum algorithm is to make superposition of many computing paths,
and then use interference to select the correct one quickly.

Along with the development of entanglement~\cite{HHH+09},
it was proved that, surprisingly, entanglement is not sufficient 
for quantum speedup~\cite{GFE09,BMW09} in the setting of measurement-based quantum computing~\cite{RB01}.
This was later extended to the circuit model by showing 
that a small amount of entanglement is enough to achieve universality~\cite{Nest13}.
This partly motivated people to investigate other resources,
such as quantum contextuality~\cite{Mer93,BCG+21}.
Indeed, it was claimed that quantum contextuality serves as the magic for 
quantum speedup~\cite{HWV+14} based on the model of magic-state injection~\cite{BK05},
and this paradigm has been expanded from then on;  
e.g., Refs.~\cite{PWB15,RBD+17,BBC+19,SRP+21}.
Recently a close relation between measurement-based quantum computing
and symmetry-protected topological order~\cite{GW09,CGW11,SPC11}  
has been established~\cite{Miy10,ESB+12,WSR17,SWP+17,ROW+19}, 
but a resource theory was not developed yet.

An important precursor for general quantum resource theory (QRT)~\cite{CG19} 
is the resource theory of quantum coherence~\cite{SAP17}.
Measures of coherence have been studied from various perspectives~\cite{BG06,Aberg06,NXZ+11,Wang11,Wang12,Sta14}, 
but only the resource-theoretic framework is complete.
Quantum features can now be rigorously defined in the framework of QRT.
Given a set $\C S$, a QRT over it is to identify a subset $\C F \subset \C S$ 
and the set of operations $\C O$ that preserves $\C F$,
and then the rest of $\C S$ is treated as resources.
A measure can then be defined to quantify the amount of resources
and used to characterize the conversion between resources.

QRT does not provide methods of how to find the so-called free set $\C F$
and free operations $\C O$, however.
In quantum computing,
this can be especially full-filled by methods from 
universal quantum computing models (UQCM) and also algorithms.
Namely, UQCM considers various settings: sets and operations on them,
and an efficient algorithm can quickly realize an operation on the set,
and these settings provide the places to define the free set and free operations.

In this work, we combine methods from QRT and UQCM and find fruitful results.
We find QRT provides a way to define and classify UQCM, and identify the universal resources,
and UQCM provides the place to define universal resources and classify them.
From it, we can resolve puzzles such as the tension between entanglement and interference,
the relation between coherence and contextuality,
and we also find important new UQCM such as the model directly based on 
quantum contextuality, which is in the same family of magic-state injection. 
Below we survey some previous work and then summarize the main findings of this work. 

\subsection{Previous work}

Due to the richness of the subject,
there are many research lines that are relevant.
Here we highlight a few points that are most relevant 
for the study in this work.

\begin{figure}
    \centering
    \includegraphics[width=0.2\textwidth]{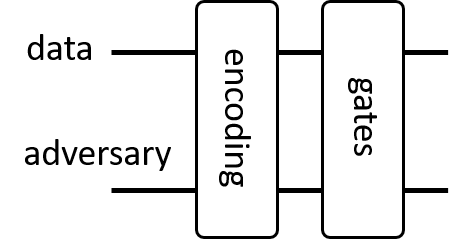}\hspace{.5cm}
    \includegraphics[width=0.17\textwidth]{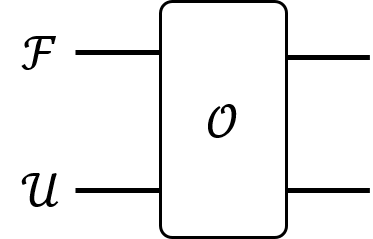}
    \caption{The ``quartet model"~\cite{Wang21} (left) and resource-theoretic model (right)
 for universal quantum computing.}
    \label{fig:ur}
\end{figure}

\begin{itemize}
    \item 
In many study of QRT, a resource is not necessarily universal,
i.e., not required to enable universal quantum computing. 
For instance, the QRT of thermodynamics defines athermality~\cite{CG19}
which does not aim for computational universality. 
Here, our study of QRT is for UQCM.
A UQCM is a framework to realize any quantum algorithms,
and two models are equivalent if tasks or algorithms in them 
can simulate each other efficiently. 
The universality requires to consider different types of operational tasks
instead of individual ones, 
as some tasks are special or interchangeable.
Therefore, we introduce \emph{universal resources} to 
characterize each UQCM.  
This provides a solid computational scheme to classify quantum resources. 
\item A classification for UQCM was developed~\cite{Wang21}.
It proposed a table for UQCM with two categories based on a ``quartet model":
with bipartite input structure and bipartite evolution structure.
See Figure~\ref{fig:ur} (left).
One category is for fault-tolerance, i.e., coding-based models, 
and one is for universality. 
In this work, we only study the category for universality,
and all the three families we identify belong to this category.
Using QRT improves the quartet model:
the bipartite input is specified to be free states and universal resources,
and the evolution gates is specified to be free operations.
Previously, measurement-based quantum computing
(MBQC) and adiabatic quantum computing (AQC)~\cite{AL18} 
were treated as coding-based models since they can be viewed as dynamic coding methods.
However, if we only restrict to static coding method, 
then MBQC and AQC must be treated using other methods.
In this work, we find MBQC belongs to the amplitude family,
and AQC belongs to the Hamiltonian family.
Our classification in this paper is an improvement of the previous one.
\item Although in the past MBQC was not studied along with QRT, 
many studies on this subject are precursors for a resource theory of it. 
The universality for MBQC was largely explored before the rising of QRT~\cite{NDV+07,NDM+07},
such as the measure of entanglement width.
A classical version of MBQC was defined~\cite{HWA+14}. 
As was mentioned, highly entangled states are mostly useless for this model~\cite{GFE09,BMW09}.
Instead, recently it was found that symmetry-protected topological (SPT) order is resourceful~\cite{WSR17,SWP+17,ROW+19},
and this lays the foundation for our resource theory of MBQC in this work.
Quantum contextuality~\cite{Mer93,Spe05,Spe08,AB11,BCG+21} 
has also been investigated for this model~\cite{AB09,Rau13},
but was not claimed to be the universal resource. 
In this work, we will further analyze the notion of quantum contextuality, 
and introduce a model directly based on it.
\end{itemize}

\subsection{Summary}

The basic model for our study is shown in Figure~\ref{fig:ur} (right).
It has two registers: the data $\C F$ and the resource $\C U$,
and the computation are free operations $\C O$. 
The initial states for data are free states,
so it is clear that using resource $\C U$ enables a computation
by consuming it.
In a resource theory, 
a free set $\C F$ can be a set of quantum states.
If this is not the case, we can convert it into a set of states
by considering the effects on states.

In quantum theory, we mainly study states, evolution, observable, and probability from measurement.
Therefore, we can identify four families of UQCMs from each of them.
More specifically, we consider the set of local Hamiltonian for the observable family.
So we introduce the amplitude (or state) family, 
(quasi-)probability family, Hamiltonian family, and evolution family of UQCMs.
The later one will be studied separately.
Therefore, a family of UQCM is identified by the set $\C S$.
A UQCM can be defined from a QRT, $(\C F, \C O, \C R)$.
Then depending on a hierarchy of resource theories, 
i.e., a subset hierarchy of a few free sets, 
we can define the generations in a family of UQCM.
Here we identify three generations for each family.
We must be aware that such generations are neither unique nor complete
due to the flexibility of the hierarchy of resource theory. 

We now briefly describe the contents of the three families:
denoted as the $a$, $p$, and $h$ families.
The $a$-family contains the quantum circuit model (QCM),
local quantum Turing machine (LQTM)~\cite{Wang20T},
and MBQC. 
In QCM, for quantum algorithmic speedup or primacy, 
we identify quantum interference as the resource. 
For LQTM, the ebit (or Bell state) is the universal resource,
serving as quantum memory and enabling the 
representation of any states as matrix-product states (MPS)~\cite{AKLT87,FNW92,PVW+07}.
In MBQC, a resource MPS is measured by a sequence of adaptive on-site projective measurements.
We identify a special feature relating to SPT order as its universal resource. 

The $p$-family contains the contextual quantum computing (CQC) that we introduce,
magic-state injection (MSI), and post-magical quantum computing (PMQC) that is motivated by 
a nonlocal teleportation scheme~\cite{Bro16}.
The MSI is powered by the so-called magic, 
often held by the $|t\ket=T|+\ket$ state.
The magic is stronger than contextuality, though,
which then allows a UQCM directly based on contextuality.
It turns out this CQC model is interesting, 
and can explain features of some quantum algorithms,
such as the linear combination of unitary operations~\cite{CW12,Long11}.
Using even stronger nonlocal boxes~\cite{PR94,Bro16},
the communication complexity can be brought down to minimal, 
enabling the PMQC model which is a variation of blind MBQC~\cite{BFK09}.

The $h$-family contains Hamiltonian quantum simulation 
(HQS)~\cite{CMP18,KPB+20,KPB+21,ZA21},
Hamiltonian quantum cellular automata (HQCA)~\cite{Arr19,Far20,Wie08},
and AQC~\cite{AL18}. 
It relies on the set of local Hamiltonian interaction terms 
and how to manipulate them.
The HQS allows almost any forms, 
HQCA allows the so-called parallel ones,
and AQC only allows adiabatic changes of interaction terms.
It is then interesting to see that 
they indeed form a family and there is a hierarchy of interaction effects.

Physically,
the $a$, $p$, and $h$ families are distinct and their major physical 
characters can be understood from the perspective of
quantum coherence, correlation, and interaction, respectively. 
There is indeed a hierarchy of resources for each of them.
For states, this is from coherence to entanglement, and to a special multipartite entanglement.
For probability, this is from local correlations of a few bits to nonlocal correlations. 
For Hamiltonian, this is also from local to nonlocal ones.
A more powerful resource can be prepared from a less powerful one
by proper operations defined from their free operations.
A hierarchy is precisely characterized by a family of resource theories.

For clarity, the main findings are listed below:
\begin{itemize}
    \item We extend resource theory to the setting of universal resources.
    This finds a place to use resource theory, 
    and draw connections between quantum information and quantum computing.
    \item We use QRT to study UQCM and introduce families of UQCMs,
    which stands as a unique approach to unify
    and classify them, examine their physics, and find new ones.
    \item We define a universal resource theory for MBQC,
    and we clarify the relation with contextuality.
    \item We introduce the models of CQC and PMQC, as generations in the $p$-family.
    \item We present a primary resource-theoretic study for Hamiltonian-based models.
    \item We study the relation between quantum algorithms and resources.
    This helps to resolve puzzles, inspire ideas for new quantum algorithms. 
\end{itemize}

Besides, we leave the study of the evolution family to the future,
which is relatively less well studied.
It may relate to superchannels~\cite{CDP08a}, dynamical resource theory~\cite{CG19}, 
and quantum von Neumann architecture~\cite{Wang22}.
Also we believe there should be at least one coding-based family, 
which contains models based on error-correction codes~\cite{Wang21}.
This would include topological quantum computing~\cite{NSS+08}, 
multiparticle quantum walk~\cite{CGW13}, for instance.
Also note that our classification of UQCMs does not aim to 
cover all known existing models.
Instead, it provides a resource-theoretic framework to investigate and fertilize them. 

This work contains the following parts.
In Section~\ref{sec:ur}, we review QRT and then define the universal resources.
We then introduce families of UQCMs.
We study the details of each family in Sections~\ref{sec:a-family},
\ref{sec:p-family}, and \ref{sec:h-family}.
We then analyze a few quantum algorithms in Section~\ref{sec:alg}.
We conclude in Section~\ref{sec:conc} with perspectives of future directions.
For convenience, the various abbreviations for the models studied in this work
are summarized in Table~\ref{tab:abbr}. 

\begin{table}[b!]
    \centering
    \begin{tabular}{|l|l|} \hline
       QCM  &  quantum circuit model \\ \hline
       LQTM  & local quantum Turing machine \\ \hline
       MBQC  & measurement-based quantum computing \\ \hline
       CQC  &  contextual quantum computing \\ \hline
       MSI  &  magic-state injection \\ \hline
       PMQC  & post-magical quantum computing \\ \hline
       HQS  &  Hamiltonian quantum simulation \\ \hline
       HQCA  & Hamiltonian quantum cellular automata \\ \hline
       AQC  &  adiabatic quantum computing \\ \hline
    \end{tabular}
    \caption{The abbreviations of models and their full terms.}
    \label{tab:abbr}
\end{table}

\section{Universal Resource}
\label{sec:ur}


\subsection{Resource}

We start from a brief review of QRT over a set of states. 
We consider finite-dimensional Hilbert spaces.
For a quantum system with a Hilbert space $\C H$ of pure states and the set of density operators $\C D$,
a resource theory is defined by a tuple 
\be (\C F, \C O, \C R) \ee 
with $\C F \subset \C D$ as the set of free states,
$\C O: \C F \ra \C F$ the set of free operations, 
and $\C R := \C D \backslash \C F$ the set of resource states~\cite{CG19}.
To quantify resources, we require the following axioms for a measure $f$:
\begin{itemize}
    \item[i)] It is zero for free objects; $f(\rho=0)$, $\forall \rho \in \C F$;
    \item[ii)] It is positively upper bounded for finite dimensional Hilbert spaces;
    \item[iii)] It is asymptotically continuous; $f(\rho)\ra f(\sigma)$ whenever $\rho \ra \sigma$, 
$\forall \rho, \sigma \in \C D$;
\item[iv)] It is subadditive;  $f(\rho\otimes \sigma)\leq f(\rho)+f(\sigma)$, 
$\forall \rho, \sigma \in \C D$;
\item[v)] It is non-increasing under free operations; $f(\rho)\geq f(\Phi(\rho)),$  
$\forall \Phi \in \C O$.
\end{itemize}

The subadditivity condition iv) can be strengthened to additivity for some measures of resources
that we will use in this work.

There are a few generic approaches to quantify the amount of resource
of a state, $\mathscr{R}(\rho)$,
including distance-based measures, entropy, Fisher information,
witness, and majorization~\cite{CG19}.
For instance,
for a contractive distance $d$, a distance-based measure of resource is 

\be \mathscr{R}_d(\rho):= \min_{\sigma\in \C F} d(\rho, \sigma). \ee
This includes the trace distance of resource and fidelity-based measures.
The relative Renyi entropies can also be used. 
For the widely used
$S(\rho \| \sigma)=-\text{tr} [\rho(\log \sigma -\log \rho)]$,
the relative entropy of resource is 
\be \mathscr{R}_r(\rho):= \min_{\sigma\in \C F} S(\rho\| \sigma).\ee
It was shown that there exists a free parameter estimation protocol for any resource state $\rho\in \C R$
with an effective
Fisher information being nonnagative~\cite{TNR21}.

In this work, we also consider QRT more abstractly. 
The set $\C F$ does not need to be a set of states. 
Instead, it could be a set of Hermitian operators, unitary operators, 
quantum channels, or measurements, for instance. 
We can always use resource measures of states
for these cases by considering operation effects on states.
Also we do not employ resource conversions such as distillation 
as our focus will be on universal resources for quantum computing. 

\subsection{Universal resource}

We now define the notion of universal resource,
which, roughly speaking, is the resource that enables
universal quantum computing. 
Formally, we define a universal resource theory as 
\be (\C F, \C O, \C R, \C U) \ee
with an additional set $\C U \subset \C R$ as 
the set of universal resource states,
compared with a usual resource theory. 
The universality means that 
$ \C O (\C F \otimes \C U) $
simulates any quantum algorithms efficiently. 
See Figure~\ref{fig:ur} (right). 
Here, efficiency means that the costs for the free operations $\C O$,
free states $\C F$, and universal resources $\C U$ 
all do not grow exponentially fast with the size of the given algorithm.

Actually, this formalism offers an unique way to define 
a universal quantum computing model (UQCM).
From computer science, there is an algorithmic approach:
for a class of problems, a family of algorithms exist to solve them with a universal way
to measure the complexity of each algorithm, e.g., circuit cost, oracle calls, or steps in Turing machines.
Two models are equivalent if algorithms from them are polynomially equivalent.

From our QRT formalism, we need to find $\C O$ and $\C F$.
The $\C F$ relates to how information is represented.
This is important since it relates to how to measure the complexity of algorithms.
With QRT, the cost of $\C O$ and $\C F$ shall be counted separately from the cost of $\C U$,
with the later being more important to measure the complexity of algorithms.
This is an improvement of the quartet model we introduced before~\cite{Wang21},
Figure~\ref{fig:ur} (left).
The adversary now serves as the universal resource,
and the data register is the free states.
The gates are free operations $\C O$,
which can be logical, i.e., an encoding may be implicit. 

A universal resource is the optimal resource defined in a resource theory
in order to prompt $\C O$ to achieve universality.
Its value is given by
\be \mathscr{R}(\C U):= \max_{\rho\in \C U} \mathscr{R}(\rho). \ee
This formula can be relaxed to be 
$\mathscr{R}(\C U)= \max_{\rho\in \C R} \mathscr{R}(\rho)$
since most likely $\mathscr{R}(\rho)$ would be the same for all $\rho\in \C U$.
We see that a measure for a universal resource is not for states,
instead, it is for a set. 
For instance, it can be a measure of distance between sets
\be \mathscr{R}(\C U)= \max_{\rho\in \C R} \min_{\sigma\in \C F}  d(\rho, \sigma). \ee
In general, the $\min$ and $\max$ functions do not commute,
but for convex sets and convex measures, the von Neumann-Fan minimax theorem
shall apply to make them commute.

Furthermore, it is viable to introduce a weaker notion, ``distilled universality,''
to capture resources that can be efficiently distilled to a universal resource.
For instance, we find ebit is a universal resource,
and those entangled states that can be used to distill ebits efficiently
will be considered distillable universal resource. 
In most QRTs, it has been shown that~\cite{BG15}, asymptotically, 
all resource states become reversible using operations 
that are resource non-generating for the set of free states.
Resource conversion via distillation or other means is an important part 
of a QRT. 
In this work, we employ the universality 
that does not rely on resource distillation.


\begin{table}[t!]
    \centering
    \begin{tabular}{|c|c|c|c|c|c|}\hline
         & QCM & & LQTM & & MBQC \\ \hline
    $\C F$ & BIT & $\supset$ & SEP & $\supset$ & PRO \\ \hline
    $\C O$ & CC & $\supset$ & SLOCC  & $\supset$ & 1O1C \\  \hline
    $\C U$ & COH & $\prec$ & EBIT & $\prec$ & UENT  \\  \hline
    \end{tabular}
    \caption{The triplet of the amplitude family of UQCMs.}\vspace{0.5cm}
    \label{tab:triplet_amp}
    \centering
    \begin{tabular}{|c|c|c|c|c|c|}\hline
         & CQC &           & MSI   &           & PMQC  \\ \hline
    $\C F$ & BIT   & $\supset$ & STAB   & $\supset$ & 1STAB  \\ \hline
    $\C O$ & CC   & $\supset$ & CLIF  & $\supset$ & 1CLIF  \\ \hline
    $\C U$ & CONT & $\prec$       & MAGIC   & $\prec$       & PMAGIC  \\ \hline
    \end{tabular}
    \caption{The triplet of the probability family of UQCMs.}\vspace{0.5cm}
    \label{tab:triplet_p}
    \centering
    \begin{tabular}{|c|c|c|c|c|c|}\hline
         & HQS & & HQCA & & AQC \\ \hline
    $\C F$ & STOQ & $\supset$ & BIT & $\supset$ & PRO \\ \hline
    $\C O$ & LINEAR & $\supset$ & PARALLEL  & $\supset$ & GAPPED \\  \hline
    $\C U$ & NSTOQ & $\prec$ & COH & $\prec$ & 1DQW  \\  \hline
    \end{tabular}
    \caption{The triplet of the Hamiltonian family of UQCMs.}
    \label{tab:triplet_h}
\end{table}

\subsection{Families of models}

By considering different types of sets and operations, 
we can define different UQCMs. 
For the sets of states, (quasi-)probabilities, and local Hamiltonian,
we introduce a family of UQCMs for each of them.  
We use the term ``family'' since for each of them
there are also different UQCMs.
In particular, for each family we identify a triplet of models,
forming a hierarchy of universal resources. 
Namely, there is an order for their computing power of the universal resources.
For two resource theories, if $\C F_1 \subset \C F_2$, then $\C O_1 \subset \C O_2$.
When the total state space is fixed, 
$\C R_1 \supset \C R_2$, i.e.,
more states are treated as resourceful in the first theory.
However, to achieve universality, more resource power is needed for the first theory.
We denote 
\be \C U_1 \succ \C U_2, \ee 
which reads 
``the universal resource $\C U_1$ is computationally more powerful 
than the universal resource $\C U_2$.'' 
Then the following conversions between the two universal resources are possible
\be (\C F_2 \backslash \C F_1 ) |u_2\ket =|u_1\ket, \; \C F_1 |u_1\ket =|u_2\ket, \ee 
for universal resource states $|u_{1,2}\ket\in \C U_{1,2}$, modular free states.


We now describe the conversion between universal resources for each family, 
with the establishment of QRT for each model 
explained in the following sections. 
See the tables~\ref{tab:triplet_amp}, \ref{tab:triplet_p}, \ref{tab:triplet_h}
for a brief summary.

The $a$-family relies on the set of states. 
Information is carried by the amplitudes $\psi_i$ of pure states 
\be |\psi\ket=\sum_i \psi_i |i\ket \ee
expanded in an orthonormal basis $\{|i\ket\}$,
and computation is the arithmetic of amplitudes, i.e., interference.
A mixed state or density operator $\rho$ can be viewed as a mixture of pure states
$\rho=\sum_\alpha p_\alpha |\psi_\alpha\ket \bra \psi_\alpha|$
with a probability distribution $p_\alpha$
over a set of pure states $\{|\psi_\alpha\ket\}$.

We find the QCM, LQTM, and MBQC form a triplet of generations in the $a$-family.
Their universal resources are coherence (COH), ebits (EBIT), 
and special entanglement denoted as UENT.
From a qudit `plus' state $|+\ket=\frac{1}{\sqrt{d}} \sum_i |i\ket$,
which has maximal COH, 
an ebit $|\omega\ket=\frac{1}{\sqrt{d}} \sum_i |ii\ket$ can be generated as
\be |\omega\ket= \text{CX} |+\ket |0\ket \ee 
for the controlled-not gate CX, 
which is free in QCM but not in LQTM.
With ebits, any state can be expressed as an MPS~\cite{AKLT87,FNW92,PVW+07} form
\be |\psi\ket= (\otimes_n \C P_n) |\omega\ket^{\otimes n}  \label{eq:vbs}\ee 
for $\C P_n$ as local operators that are free in LQTM but not in MBQC.
This includes the 2D cluster state~\cite{RB01} and AKLT state~\cite{AKLT87} that contains UENT
as the universal resource.

The $p$-family relies on the set of probabilities or quasiprobabilities,
relating to measurement effects.
By expressing a state as
\be \rho = \vec{r} \cdot \vec{\sigma} \ee
in a Hermitian operator basis $\vec{\sigma}$,
it is characterized as a vector $\vec{r}$. 
Following Wootters~\cite{Woo87}, 
the vector satisfies $\text{sum}(\vec{r})=1$ in a proper basis.
If $\vec{r}\geq 0$, it can be viewed as a probability distribution,
while in general there are negative values. 
So a state is treated as a quasiprobability, i.e., Wigner function,
and a computation is the change of it.

The $p$-family is motivated by the MSI model~\cite{BK05}.
In this model, the free set is formed by stabilizer states (STAB),
all with positive Wigner function~\cite{Gro06}, and Clifford operations (CLIF) are free.
This selects out the magic state 
\be |t\ket=T|+\ket \ee 
as the universal resource,
for $T$ known as the T gate. 
However, not all states with positive Wigner function are stabilizer states~\cite{VFG+12}. 
This implies that below MSI, there is a more basic model 
that directly captures the negativity of Wigner function. 
As Wigner negativity is equivalent to quantum contextuality~\cite{Spe08},
we introduce the model of contextual quantum computing (CQC). 
In this model, we use the superposition of quantum contexts as universal resource, 
requiring states like $|t\ket$, $|+\ket$, and ebit $|\omega\ket$, 
and also measurement feedback (classical communication). 


For both models, classical communication is required to 
deterministically realize a gate. 
This reveals the role of correlations. 
There is a stronger form of correlations, 
known as Popescu-Rohrlich (PR) nonlocal box~\cite{PR94}.
It is discovered that the PR box can substitute the classical communication
for T gate teleportation~\cite{Bro16}.
This motivates our definition of the PMQC model,
which relates to MBQC and also instantaneous nonlocal quantum computation.
Quantum teleportation is free in MSI but not in PMQC.


The $h$-family is based on the set of interactions, i.e., Hamiltonian terms.
We consider local terms and the switching on and off of them as operations.
An initial input state of an algorithm can be treated as an eigenstate of a Hamiltonian 
\be H |\psi\ket= E |\psi\ket, \ee 
and then use arithmetic of Hamiltonian to change $H$.
We find HQS, HQCA, AQC form the triplet generations in the $h$-family,
with different free sets of arithmetic of interaction terms.

Although it dates back to the origin of quantum computing~\cite{Fey82,Llo96,WRJ+02,DNB+02},
we find the foundation of HQS is the notion of a universal 
set of H terms~\cite{CMP18,KPB+20,KPB+21},
just like a universal gate set.
Given a set $\C S$, a Hamiltonian is constructed by a real-weighted sum 
\be H=\sum_n j_n h_n \ee
for amplitudes $j_n\in \B R$ and each term $h_n \in \C S$. 
The evolution $U=e^{iHt}$ will be Trotterized with a sequence of local terms $e^{i t_n j_n h_n}$~\cite{Llo96}.
It has been known that almost all two-local terms are universal,
with stoquastic or classical ones as special forms~\cite{CMP18,KPB+20,KPB+21}.
We define stoquastic Hamiltonian as the free set.

The HQCA model, as a class of QCA, relies on parallel switching on and off local terms.
Each step has a special $j_n$ and also $t_n$. 
In other words, HQCA is a special HQS,
just like the case with LQTM and QCM.
It requires a special local term with larger locality for universality.
The free set is the classical CA, which can be classically universal.
Finally, the AQC uses adiabatic sum of local terms, 
with no gap-closing which protects the information encoded in the ground state manifold.
The free set is product of local states, equivalent to on-site Hamiltonian,
and the universal resource is equivalent to a 1D quantum walk derived from
the Feynman-Kitaev circuit-to-Hamiltonian map~\cite{KSV02}.


\section{The $a$-family} 
\label{sec:a-family}

\subsection{Quantum circuit model}
\label{subsec:qcm}

In this section, we recall the quantum circuit model (QCM) and draw its connection
with coherence and interference.
For simplicity, we focus on the qubit case, 
and the extension to the qudit case is straightforward.
It contains three basic stages:
\begin{itemize}
    \item[i)]  Initialize at the all-zero state $|00\cdots 0\ket$ of qubits;
    \item[ii)] Apply a sequence of unitary gates;
    \item[iii)] Readout by measurements in the Pauli Z basis.
\end{itemize}
A quantum algorithm in the QCM is often described as a sequence of gates,
possibly with additional classical pre and post processing. 
A central result is the existence of universal gate set
from which any unitary evolution in the group $SU(2^n)$ can be efficiently constructed~\cite{NC00}.
Two well-known sets are $\{$H, T, CX$\}$ and $\{$ H, CCX$\}$,
for the Hadamard gate H and T gate as
\be \text{H}=\frac{1}{\sqrt{2}}\begin{pmatrix} 1 & 1 \\ 1 & -1 \end{pmatrix}, 
\text{T}=\begin{pmatrix} \tau & 0 \\ 0 & \tau^* \end{pmatrix}, \ee 
and the controlled-not gate CX, Toffoli gate CCX,
for $\tau\equiv e^{i\pi/8}$.
The Toffoli gate itself is universal for classical computation,
which motivates our resource theory for QCM.


We now introduce the QRT for QCM.
The set of free states $\C F$ 
is the set of all classically efficiently representable states.
The set of free operations $\C O$ is all efficient classical algorithms.
That is, the free states and free operations form the usual classical computation
that solves problems in the complexity class BPP (bounded-error probabilistic polynomial). 
We denote them as BIT and CC, for convenience (see Table~\ref{tab:triplet_amp}). 
The universal resource $\C U$ will boost BPP to BQP (bounded-error quantum polynomial), 
enabling universal quantum computing and
serving as the necessary and sufficient resource for quantum speedup. 

\begin{figure}[t!]
    \centering
    \includegraphics[width=0.28\textwidth]{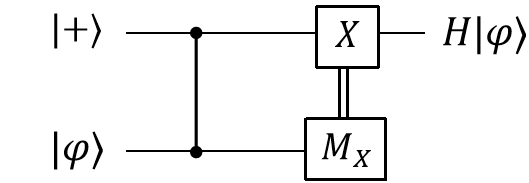}
    \caption{Quantum teleportation of H gate.}
    \label{fig:H_gate}
\end{figure}

The universal resource $\C U$ contains the plus state
\be |+\ket= \frac{1}{\sqrt{2}} (|0\ket+|1\ket)\ee 
and any finite tensor-product $|+\ket^{\otimes n}$, 
together with all their equivalents under free operations,
such as the `minus' state $|-\ket= \frac{1}{\sqrt{2}} (|0\ket-|1\ket)$. 
The plus state $|+\ket$ is the magic state for H gate
since given $|+\ket$, H can be realized by quantum teleportation,
which are free operations. 
See Figure~\ref{fig:H_gate}.
An intriguing fact is the measurement must be in the X basis,
instead of Z basis.
From a different perspective, 
we will show that this relates to the resource theory of contextuality 
studied in the next section. 

It is clear that a classical circuit with Toffoli gates 
can only change basis states among themselves, 
while H can generate arithmetic on the amplitudes,
which is interference. 
All gates that do not generate superposition are also classical,
such as CX, phase gate S, and T gate.
Free measurements not only contain Z-basis measurements, 
but also Pauli measurements.
The classical measurement outcomes can be used as classical control
over conditional classical gates,
which also form free operations.  



In QCM, a quantum circuit is usually not expressed 
with injection of plus states. 
Instead, it is often a sequence of unitary gates. 
In order to measure the power of quantum circuits, 
the QRT of coherence (COH) can be applied but it is defined for states.
We introduce a new measure of interference (INT) power for gates,
serving as the coherence measure of gates.


Given a state $\rho$, the $\ell_1$-norm measure of coherence is 
\be C(\rho)=\sum_{i\neq j} |\rho_{ij}|. \ee
It is not additive, though.
It can be converted into an additive measure
\be Q(\rho)=\log_2(C(\rho)+1), \ee
and $C(\rho)+1$ is the $\ell_1$-norm of $\rho$.
We call $Q$ as the logarithmic coherence, or ``log-coherence'' for short. 
It is clear that
\be Q(\rho_1\otimes \rho_2)=Q(\rho_1)+Q(\rho_2).\ee
It's maximum is $\log d$, for $d$ as the dimension,
and this agrees with the maximum of the relative entropy of coherence
\be C_r(\rho)=S(\Delta(\rho))-S(\rho)=\min_{\sigma\in \C F} S(\rho\| \sigma),\ee
which is additive~\cite{WY16}, 
for $S$ as von Neumann entropy, $\Delta$ as the completely dephasing channel.
With the eigenvalues $p_i$ of a state $\rho$,
$S(\rho)$ is the Shannon entropy $H(p_i)$ of the probability distribution $p_i$.


To quantify interference, we need a dynamic measure of coherence.
From channel-state duality~\cite{Jam72,Cho75}, a process $\C E$ is mapped to a Choi state
\be \C E \mapsto \omega_{\C E}:= \I  \otimes \C E (\omega)= \frac{1}{d}\sum_{ij} |i\ket\bra j| \otimes \C E(|i\ket\bra j|), \ee
for the ebit $\omega=|\omega\ket\bra \omega|$.
Note we act the channel on the second part of the ebit for convenience. 
Then the coherence of this state can be treated as the interference power of the gate.
However, this does not work.
For instance, the Pauli gate $X$, $Y$, $Z$ each is mapped to a Bell state
which has nonzero coherence,
yet we expect the interference power of Pauli gates is zero.
This difficulty can be resolved by modifying the duality based on the following observation:
the domain of such a measure is the free set, instead of the whole state space.
We introduce a classical channel-state duality as follows.
\begin{definition}[Classical channel-state duality]
For a quantum channel $\C E$, its classical dual state is defined as 
\be \C E \mapsto m_{\C E}:= \I  \otimes \C E \circ \Delta (\omega)= \frac{1}{d}\sum_i P_i \otimes \C E(P_i), \ee
for projectors $P_i=|i\ket\bra i|$, $\Delta$ as the completely dephasing channel.
\end{definition}
The dephased Bell state $\I\otimes \Delta (\omega)$ is maximally classically correlated.
The coherence of the state $m_{\C E}$ is
\be C(m_{\C E})=\frac{1}{d} \sum_{i} C(\C E(P_i)), \ee
and the log-coherence is additive 
\be  Q(m_{\C E_1\otimes \C E_2})= Q(m_{\C E_1} \otimes m_{\C E_2})= Q(m_{\C E_1})+ Q(m_{\C E_2}). \ee
The relative entropy of coherence is
\be C_r(m_{\C E})= \frac{1}{d} \sum_{i} C_r(\C E(P_i)),\ee
and it is indeed additive.
From the relative entropy and $\ell_1$-norm,
the interference power of a gate $\C E$ can be viewed as the average of the 
coherence of the output states $\C E(P_i)$ each created from a free state $P_i$.
We can now apply our measure to convince the intuition that
Pauli gates have zero interference power,
while Hadamard gate has maximal interference power on a qubit.
We will apply this to study quantum algorithms in Section~\ref{sec:alg}.


\subsection{Local quantum Turing machine}
\label{subsec:lqtm}

We now consider a computing model for which the free set of states
is smaller than the set of classical states.
The QRT of entanglement (ENT) is a natural fit. 
For the bipartite case,
a state is separable if it can be written as 
\be \rho= \sum_i p_i \rho_i^A \otimes \rho_i^B \ee
for bipartite partition $A|B$ of the system.
We also require that local dimensions are constants. 
The free operation $\C O$ is 
stochastic local operation and classical communication (SLOCC),
which can only preserve or decrease ENT. 
For a bipartite setting $\C H_d \otimes \C H_d$,
states $\frac{1}{\sqrt{d}} \sum_i e^{i\theta_i}|ii\ket$ are maximally entangled
with ENT of $\log d$, including Bell states.

We find the computing model that relies on ENT as resources
is the LQTM~\cite{Wang20T}, as a simplification of the original ones~\cite{BV97,Yao93,MW19}.
In this model, a computation requires two registers:
one for the data, and the other as an adversary or ``machine.''
The machine interacts with each qubit in the data register one at a time,
preparing a MPS 
\be |\psi\ket= \sum_{i_1,\dots,i_N} \text{tr} (B A^{i_N} \cdots A^{i_1} ) |i_1\dots i_N\rangle,\ee
for a boundary operator $B$,
whose role has been analyzed in details~\cite{Wang20T}.
The data qubits do not interact with each other directly,
which is a natural locality constraint,
and suggests ENT as the resource for LQTM.

The set of matrices $\{A^{i_n}\}$ at each local site $n\in \{1,2,\dots,N\}$ form a channel $\mathcal{E}_n$
acting on the adversary space, 
which in other contexts may be known as
the virtual, bond, or edge space, or correlator.
The dimension of the adversary identifies the amount of ENT for the whole system.
As each $\mathcal{E}_n$ can be dilated to a unitary operator $U_n$, 
the MPS can be prepared by a sequential circuit. 


Another way to represent MPS, actually the original one,
is the VBS method~\cite{AKLT87}.
Note that the term MPS is often used for 1D systems with small ENT,
while tensor-network states (TNS) or PEPS are for higher dimensions,
but mathematically, they are all MPS~\cite{Sch11}.
Given a collection of ebits, applying operators $\C P_n$ as SLOCC 
will prepare the MPS, see Eq.~(\ref{eq:vbs}). 
This makes the role of ebits as universal resources $\C U$ explicit. 
Non-maximally entangled states require distillation scheme~\cite{HHH+09},
therefore are weaker than ebits.
Without ebits, the SLOCC, as free operations $\C O$, 
only generates separable states (SEP), as the free set $\C F$. 


Although there are local coherence for separable states,
they can be efficiently classically simulated, namely, 
by realizing the mixing and local states each with constant local dimension. 
The total Hilbert space dimension effectively does not grow exponentially with the system size.
The local coherence under SLOCC cannot lead to ENT or large amount of INT.

The relation between ENT and COH has been well studied, e.g. Refs.~\cite{SSD+15,ZMC+17}.
Given a bipartite setting $\C H_A \otimes \C H_B$ and the orthonormal 
product basis $\{|a,b\ket\}$, as the set of extreme incoherent states,
the ENT of a state is upper bounded by its COH as
\be E(\rho)\leq \min_{U} C(U\rho U^\dagger) \ee 
for $U=U_A\otimes U_B$ as a product unitary operators.   
This means we can study the ENT power of a gate
based on the INT power of it.
On the contrary, COH can also be quantified by ENT as 
\be C(\rho)=\lim_{d_a\ra \infty} \sup_\Lambda E(\Lambda (\rho \otimes |0\ket\bra 0|)) \ee
for $d_a$ as the ancilla dimension, $\Lambda$ as a bipartite incoherent operation~\cite{SSD+15}.
Using relative entropy, there exists a $\Lambda$ (a generalized CNOT gate)
that achieves the sup for $d_a\geq d$.
The ENT itself is also the COH of the final state. 

Physically, ENT and COH are not the same.
COH describes global feature of a system.
ENT is a special COH, shared or distributed,
and it identifies the quantum correlation among parts of a system.
Such a quantum correlation can be understood as quantum memory. 
In a different study, the channel derived from MPS is termed as 
channel with memory~\cite{CGL+14}.
For LQTM, it is indeed intuitive to treat the control machine as ``memory.''
In quantum communication, building up a remote ebit 
can send quantum information from one party to the other~\cite{BBC+93}.
Recently, it is shown that ebit is the basic element for quantum memory~\cite{Wang22}
relying on the channel-state duality.
A stored quantum program is a Choi state built from ebits,
the local measurements on which execute the read/write operations.






\subsection{Measurement-based quantum computing}
\label{subsec:mbqc}

\begin{figure}[b!]
    \centering
    \includegraphics[width=0.3\textwidth]{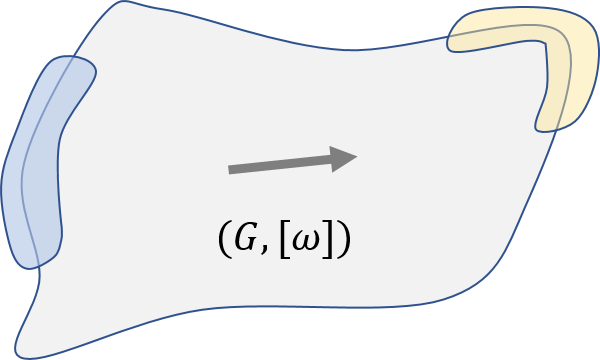}
    \caption{A schematics for the information flow (the arrow) from 
    the input (blue region) to the output (yellow region) of a computation in MBQC
    that is labelled by $(G,[\omega])$.}
    \label{fig:mbqc}
\end{figure}

We now study a model that identifies special entangled states as universal resources,
hence more restrictive free states and operations.
For instance, the Bell-state measurement~\cite{NC00} will not be a free operation.
This is MBQC, which often contains two parts: 
\begin{itemize}
    \item[i)]  A universal resource state;
    \item[ii)] A sequence of on-site adaptive measurements.
\end{itemize}
A classical side-processing of measurement outcomes is required.
The well-known original state is the 2D cluster state~\cite{RB01} 
and the underlying mechanism for gate execution is identified as gate teleportation~\cite{CLN05}. 
It was later extended to MPS by writing a resource state as 
\be |\psi(\ell)\ket= \sum_{i_1,\dots,i_N} A^{i_N} \cdots A^{i_1} |\ell\ket |i_1\dots i_N\rangle,\ee
with the data $|\ell\ket$ carried by edge space, 
and a sequence of on-site adaptive local projective measurement
(LPM) on the bulk to induce universality on it~\cite{GE07}.
The LPM can also be extended to local POVM~\cite{WAR11}.
Recently, this is also understood as a one-way code switching on edge codes~\cite{Wang21}.
The switching can be made fault-tolerant by code concatenation and error correction.





Without the resource state,
on-site adaptive measurements can only generate product states.
So we identify the set of free states $\C F$ as product states (PRO), 
which is a subset of SEP.
The free operations $\C O$ are 1-site operations (1O) 
and 1-way classical communication (1C), denoted as 1O1C. 
The 1O1C acting on PRO does not generate SEP, in particular, 
it does not generate globally shared pbits.
This type of pbits shall be generated by 1O1C on the resource state.

Recently, the close connection between MBQC and SPT order is shown~\cite{WSR17,SWP+17,ROW+19}.
SPT ground states are injective and have a well-defined locality.
Namely, its local sites are defined such that the local tensor $A$ is injective.
The injectivity means that we can achieve any action on the edge space 
by acting on the physical spins~\cite{SCP10,SPC11}.
Often, translational invariance is present.
An injective tensor may be obtained by blocking a few sites.
This will violate the 1O1C condition.
For MBQC, it further requires the more restrictive on-site injectivity,
and this can be provided by SPT order.

At present, three classes of SPT order are known to be universal.
This includes some 2D AKLT phase~\cite{WAR11,Miy11} with weak SPT order,
2D cluster phase~\cite{ROW+19} with 1-form SPT order, 
and some hypergraph states~\cite{MM16,GGM19} with strong SPT order.
Meanwhile, it is known that some graph states are also
universal, whose ENT is captured by the measure of entanglement width~\cite{NDM+07}
and does not need to be SPT apparently.

It is not hard to show that an injective MPS (or TNS in higher D)
can be driven to a SPT phase by free operations,
although this driving is not unique.
Given a state $|\psi\ket$, identify regions for the input, output, and bulk,
and the information flow direction.
See Figure~\ref{fig:mbqc} for an illustration. 
A pair  \be (G, [\omega]) \ee 
of a symmetry, normally specified by a group $G$, 
and SPT class labelled by an element $[\omega]$ in a group cohomology 
needs to be chosen, hence fixing its local tensor $A$.
Each local tensor of $|\psi\ket$ can be modified by free local operations 
to the tensor $A$. 
In other words, an injective TNS can be viewed as a SPT state with disorder and lattice defects
(e.g., a few sites might be missing).
For instance, universal graph states and weighted graph states 
can be viewed as variations of the 2D cluster state.
Furthermore, using the distillation-like technique~\cite{SWP+17},
states in a SPT phase can be driven to its fixed point by LOCC.

Therefore, we identify fixed points of 2D SPT order as universal resources.
More precisely, any fixed point of 2D SPT order that is equivalent to the 2D cluster state under 
the 1O1C free operations. 
The notion of SPT ENT has been studied~\cite{Mar17} for 1D abelian cases,
and can also be extended to higher dimensions.
The entanglement width~\cite{NDM+07} shows that 
the ENT of universal resources shall scale linearly with the logical system size.
Therefore, we use the term universal-ENT (UENT) to identify 
the ENT in a universal resource for MBQC.

For resource conversion, it is easy to see from the MPS form,
the 2D cluster state (and its equivalents), 
is prepared by applying LOCC that is not 1O1C on ebits.
The local operations~(\ref{eq:vbs}) in MPS are precisely of this nature. 








It has been proven that random states with large ENT is not useful for MBQC~\cite{GFE09,BMW09}.
Indeed, the highly entangled states under MBQC mostly behaves like local pbits.
So there is no globally shared pbits.
This means universality requires some global features of the resource states,
and this relates to symmetry which is indeed a global feature of quantum systems. 
For universality, we need both extensive ENT and on-site injectivity,
the former is provided by the gapped feature of 2D systems,
and the later is provided by the feature of SPT order. 
We shall also note that this does not mean highly entangled random states 
are useless in some other models.


\begin{figure}[t!]
    \centering
    \includegraphics[width=0.35\textwidth]{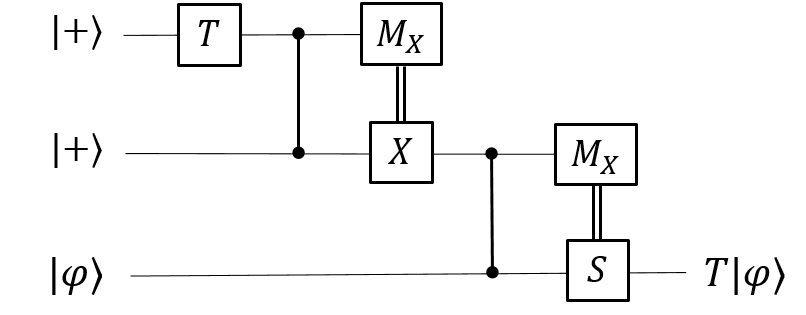}\vspace{.5cm}\\
    \includegraphics[width=0.35\textwidth]{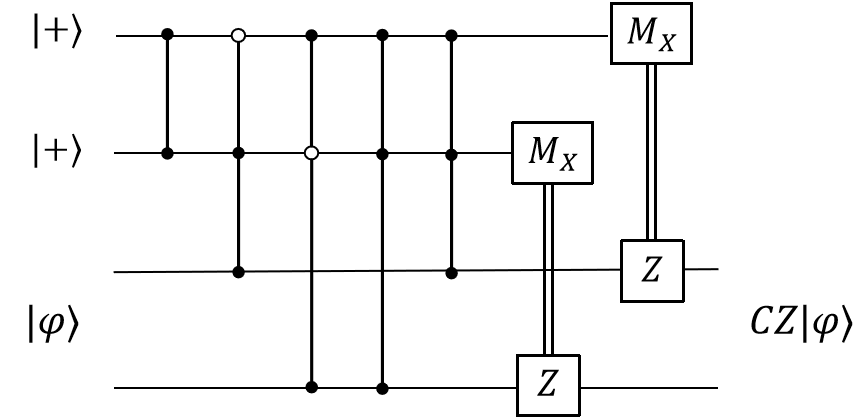}
    \caption{Quantum circuits to realized the T gate and CZ gate.}
    \label{fig:cont_gate}
\end{figure}

\section{The $p$-family} 
\label{sec:p-family}

\subsection{Contextual quantum computing}
\label{subsec:cont}

In this section, we introduce the model of contextual quantum computing (CQC). 
We start from the physics of contextuality,
which has an origin from the hidden variable theories~\cite{Mer93}.
A quantum context is a quantum operation, e.g., state preparation, evolution,
or measurement~\cite{Spe05}. 
When two operations commute, then they are compatible, so can exist simultaneously.
A compatible setting does not have contextuality, meaning that 
operators can be reduced to numbers. 
Quantum contextuality refers to the simultaneous existence of incompatible quantum contexts.
Classical contextuality can be defined as
the mixture of incompatible quantum contexts.
We then characterize quantum contextuality (CONT) as superposition of quantum contexts.
This motivates the model of CQC.

By expressing in the Pauli basis, 
we find the circuits for T and CZ in Figure~\ref{fig:cont_gate},
also see the Figure~\ref{fig:H_gate} for H gate.
If Pauli operators are treated as primary quantum contexts,
then each gate shows quantum CONT. 
These contextual circuits are universal since H, T, CZ form a universal gate set.

\begin{figure}[t!]
    \centering
    \includegraphics[width=0.2\textwidth]{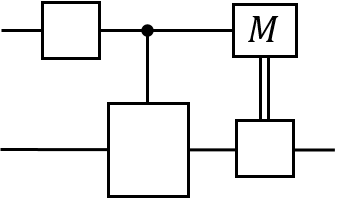}\hspace{1cm}
    \caption{A schematics of the contextual quantum circuit.}
    \label{fig:concir}
\end{figure}

We now introduce a general definition of contextual quantum circuit,
with an illustration in Figure~\ref{fig:concir}.
\begin{definition}
[Contextual quantum circuit] \label{def:contcir}
A contextual quantum circuit contains two registers: one as control, one as data,
and it realizes process of the form
\be \emph{tr}_c \circ (CV (U_2\otimes \I) CU (U_1\otimes \I)) \ee 
with a sandwiched structure: 
a special initial control state prepared by $U_1$, 
a special measurement on the control prepared by $U_2$ with 
feedback to the data register, realized by $CV$ and the trace over control,
and the quantum-controlled gates $CU$ in the middle with the data as the target.
The two controlled gates, also known as multiplexer, take
the form \be CU=\sum_i P_i \otimes U_i, \label{eq:plex}\ee
for projectors $P_i=|i\ket\bra i|$ on the control,
and unitary $U_i$ on the data.
\end{definition}
It realizes a gate $U$ deterministically
by expressing it as a linear combination of gates. 
The quantum control register is necessary since without it,
a directly applied gate only leads to superposition of states.
A mixing of contexts is realized by classical control.

Our definition can be slightly extended by using more general gates in the middle,
which would contain feedback action on the control.
We would not pursue this in details here. 

In this model, measurements are important.
Quantum measurement is described by POVM.
A POVM is a set $\{E_i\}$ for effects $E_i\geq0$ and $\sum_i E_i=\I$.
A measurement on a state $\rho$ yields three pieces of information:
the outcome $i$, the probability of outcomes $p_i=\text{tr}(E_i\rho)$,
and the final state $\rho_i$ for each $i$.
It is selective if the index $i$ is explicitly known,
and non-destructive if $\rho_i$ is available.
If it is a mixture, then it is effectively a quantum channel. 

The classical outcome $i$ is crucial to introduce in measurement-controlled operations:
depending on $i$, different operators can be further applied. 
This is crucial for quantum teleportation and also for MBQC:
the Pauli byproduct correction needs to be conditioned on previous measurement outcomes. 
Actually, contextual circuits can be seen as extensions of quantum teleportation.


We can now define the QRT of CQC. 
The free set is for BIT and CC, just like QCM,
with the input for an algorithm as bits or pbits,
computation circuit formed with S, T, CX, CZ, CCX, and any diagonal gates,
but readout measurement only in the Z basis,
and also Z-measurement controlled circuits.
Such circuits can never generate superposition of states. 

It is not hard to see the universal resource is the Pauli measurement $M_X$. 
The selective $M_X$ not only prepares the initial resource state $|+\ket$,
but also lead to the CONT. 
With free CZ and T gates, it also generates $|\omega\ket$ and $|t\ket$.

Therefore, in our framework CONT is equivalent to INT.
For QCM, we treat $|+\ket$ as the resource but $M_X$ as a free operation.
However, strictly speaking, $M_X$ can prepare $|+\ket$ 
and is more suitable to be viewed as the resource-equivalence of Hadamard gate.  
We can use COH and INT directly as the measures for CONT of states and gates. 
For instance, the CONT of $|t\ket=T|+\ket$ is 1, 
the value of COH of $|+\ket$. 
We will study the relation between CONT and INT more in Section~\ref{sec:alg}. 


Although it is far from complete,
the study above lays the foundation of CQC,
At this point, it is interesting to draw the connections with other models.
First, compared with QCM,
the role of quantum control is explicit in CQC. 
A generic contextual circuit can have many control registers.
A potential application is the setting of modular computing
where control of unknown operations, as black boxes or oracles,  
is needed~\cite{AFC14,TMV18,GST20,VC21}. 
An example is Kitaev's quantum phase estimation algorithm~\cite{KSV02},
which can estimate the phase factor $\theta$ for an unknown $U$
but known eigenstate $|\psi\ket$ of it, 
with $U|\psi\ket=e^{i\theta}|\psi\ket$.



The idea of interference of operators was studied in the model of duality QC~\cite{Long06,Long11},
leading to the algorithm of linear combination of unitaries (LCU). 
LCU has been used in Hamiltonian-evolution simulation and others,
which in general requires post-selection on the control register~\cite{CW12,BCC+15,WL16},
making it probabilistic.
This can be avoided for special cases of LCU,
followed with measurement-controlled operations.
To simulate a gate $U$ in CQC,
it first can be decomposed as a sequence of H, T, CZ, 
and then each of them can be deterministically realized by a contextual circuit. 

Finally, CONT in MBQC has been studied~\cite{AB09,Rau13},
since measurement plays a crucial role. 
From the teleportation picture of MBQC~\cite{WSR17},
a gate is simulated by gate teleportation. 
We can treat the ancilla plus $|+\ket$ or ebit $|\omega\ket$ state 
and $M_X$ as the resource for teleportation.
The MBQC with the 2D cluster state is universal and must be contextual,
and it requires measurements away from the Z basis.
However, the teleportation picture will break a MBQC process into pieces,
losing the global feature of it. 
Instead, we find our QRT of MBQC in Sec.~\ref{subsec:mbqc} is 
more natural for UENT, 
which is first given as a universal resource, 
and then consumed by free on-site local adaptive POVM. 
On the contrary, in CQC the role of contexts is explicit,
and there is no need to prepare a highly entangled state at first.





\subsection{Magic-state injection}
\label{subsec:msi}

The resource theory for MSI has been well developed~\cite{VMG+14}.
Here we briefly describe it for completeness.
This model is suitable for fault-tolerant quantum computing with stabilizer codes,
which usually allow transversal Clifford logical gates. 
In this model, the free set $\C F$ is formed by stabilizer states (STAB)~\cite{Got98},
all with positive Wigner functions~\cite{Gro06}, 
and Clifford operations (CLIF) are free $\C O$.
This selects out the magic state $|t\ket=T|+\ket$ as the universal resource,
for $T$ known as the T gate. 

A seminal additive measure is known as the `mana'~\cite{VMG+14}.
With the standard Wigner function $W_\rho$ for odd dimensional qudit states~\cite{Woo87,Gro06}, 
the mana is \be M(\rho)=\log (2 N(\rho)+1), \ee 
for the sum negativity $N(\rho)=\sum_u'|W_\rho(u)|$ for $W_\rho(u)<0$.
To draw the connection with COH, 
it is clear that $N(\rho)\leq C(\rho)$,
as the former measures the distance from STAB, 
while the later measures the distance from an incoherent set,
as a subset of STAB~\cite{MSP18}.
This is similar with the fact that ENT is smaller than COH for a state~\cite{SSD+15}.







\subsection{Post-magical quantum computing}






What would be a UQCM for which the universal resource is more
powerful than MAGIC?
We need to identify a subset of STAB.
However, this is not unique. 
Here we study a model that relies on a stronger form of CONT than MAGIC,
which is a type of instantaneous nonlocal QC (INQC),
and has a close connection with MBQC.

We have seen that measurement feedback is useful.
On the contrary, there are settings which do not allow 
or have serious limitations on this.
An extreme example is INQC, 
which, instead of classical communication, 
only allows the broadcast of local results. 
Such class of operations has been termed as 
LOBC~\cite{GC20}, as a subset of LOCC. 
A primary setting is a two-party nonlocal task:
A and B each gets an input $x$, $y$, and 
then use LOBC operations to
output $a$, $b$, which are then used to compute the result $f(x,y)$.
Security is a natural requirement,
and here we consider the so-called one-sided security~\cite{Col07},
wherein one party can know the computation result. 




\begin{figure}[t!]
    \centering
    \includegraphics[width=0.4\textwidth]{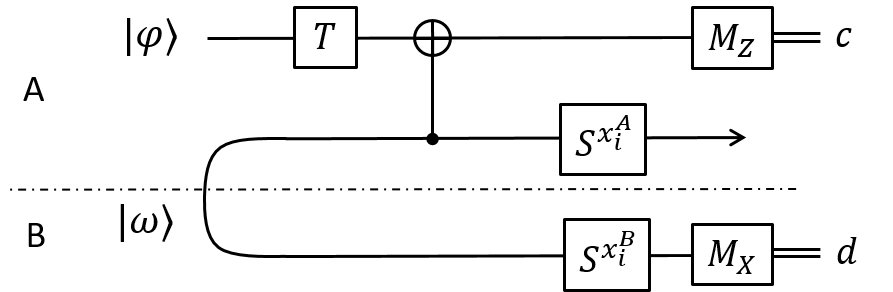}
    \caption{Broadbent's non-signaling T gate teleportation. 
    The curve is the ebit $|\omega\ket$. 
    The middle wire with arrow carries the output.}
    \label{fig:btt} 
\end{figure}

A seminal scheme for encryption is to use Pauli operators $X^a Z^b$
for encoding, with $a,b\in\{0,1\}$ as the keys~\cite{AMT+00}.
It is discovered by Broadbent~\cite{Bro16} that 
the measurement feedback for T gate teleportation can be replaced by
the Popescu-Rohrlich (PR)-box~\cite{PR94},
which on input $x$, $y$ will output $a$, $b$ satisfying
\be a \oplus b= x \cdot y, \ee
and this achieves the maximal violation of CHSH inequality~\cite{CHSH69},
larger than the quantum value.
Broadbent's nonlocal T gate teleportation (BTT),
shown in Figure~\ref{fig:btt}, forms the 
starting point of our PMQC model.
In the usual T gate teleportation, a phase gate S needs to be corrected due to 
\be T X^a Z^b = X^a Z^{a\oplus b} S^a T. \ee
This also means a T gate will destroy the update rule of the key.
The phase gate is avoided by using an ebit and a PR box.
The PR box is used to linearize $(x_i^A\oplus c)x_i^B$ as $z^A \oplus z^B$,
and the ebit is used to inject the values $x_i^A$, $x_i^B$ 
at a later time to cancel the S gate.

\begin{figure}
    \centering
    \includegraphics[width=0.2\textwidth]{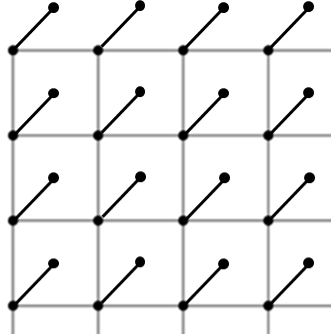}
    \caption{A schematics of a tailed cluster state $|\Phi\ket$.}
    \label{fig:tcl}
\end{figure}

We now introduce the PMQC model.
It is better described as an extension of the MBQC with the 2D cluster state. 
Instead of the usual cluster state, 
now each qubit has a `tail' (or partner), as in the Figure~\ref{fig:tcl}, 
and a tailed cluster state 
\be |\Phi\ket =\left( \bigotimes_e CZ_e \right) |\omega \ket^{\otimes n}_{AB} \ee 
is prepared as follows.
Identify all `heads' (`tails') of the collection of ebits as party A (B).
Given the ebits $|\omega \ket^{\otimes n}_{AB}$ as the initial state, 
a $CZ_e$ gate is applied between the nearest neighbor of party A 
for each edge $e$ of the square lattice.
Such a circuit is an example of tailed quantum circuits~\cite{Wang22}.

The universality of the 2D cluster state is shown by its ability to simulate 
the CZ gate and any qubit rotations, 
identifying each row as a logical evolution direction of a qubit.
Instead of arbitrary qubit rotations, we only require H and T gates,
which is universal for SU(2)~\cite{NC00}.
The H gate is realized by $M_X(a)$, leading to $HZ^a$, 
while T gate is realized by $M_X(b)TM_X(a)$, leading to $HZ^bHZ^aT=X^bZ^a T$.
To implement a sequence of H and T on encrypted state in the PMQC model, 
see the top row of the cluster,  
the input encrypted state $X^aZ^b |\psi\ket$ is injected by measuring the leftmost tail site.
The Pauli byproduct in a sequence like $HX^aZ^b T H X^cZ^d H T \cdots $ 
is brought out to the end via the BTT (Figure~\ref{fig:btt}).
Namely, for each site in party A, an ebit and PR box are consumed.   
A CZ gate is realized as in the original MBQC.
Therefore, we show that the PMQC model can realize the universal gate set $\{$H,T, CZ$\}$
on encrypted data, 
provided by a client party B to a server party A 
who does not know the input and output of the data, 
with only a final round of broadcast communication between A and B.



To formulate a QRT of PMQC, 
first notice that all the T gates can be applied first before the measurements.
Namely, in order to simulate a circuit of H, T, CZ gates, 
each simulated gate can be `imprinted' to a site or edge of the sub lattice A,
and T gates, which commute with CZ gates, can be applied on proper A sites  
leading to a circuit-specific cluster.
Then only Pauli-basis measurements are required to simulate the circuit. 
So we identify tensor-product of single-qubit Clifford operations,
and broadcast communication, as the free operation set $\C O$,
denoted as 1CLIF.
The free state set $\C F$, 1STAB, contains tensor-product of Pauli eigenstates as extreme points.
They are smaller than the free setting for MBQC, and also MSI. 
This selects out the circuit-specific tailed cluster states, 
and PR box together as the universal resource,
with the PR box playing the central role.
At present, we coin the term `post-magic' (PMAGIC) as the universal resource in the PMQC model.

The PR box was motivated by the study of quantum nonlocality (NONL),
which now is often treated as a special type of CONT~\cite{AB11}. 
Different from a generic setting of CONT,
NONL does not allow conditional operations and nonlocal operations. 
It has been proposed to use LOSR (pre-shared randomness) as
the free operations defining NONL~\cite{Bus12}. 
With local states as filters,
it shows that ENT is equivalent to NONL. 
At this point, it could be enlightening to compare NONL with MAGIC.
We established ENT, therefore probably NONL, 
as the resource for LQTM, sitting in between QCM and MBQC.
Magic is the resource for MSI, sitting in between CQC and PMQC.
NONL is not comparable to MAGIC, 
since MAGIC relies on conditional operations but NONL does not. 
MAGIC also depends on the Clifford hierarchy~\cite{GC99}. 
We see that PMQC is a combination of MAGIC and NONL.




It's quite surprising the PR box is needed to achieve universality
since its correlation is beyond quantum theory. 
Also the PR box does not need to be perfect:
a slight post-quantum correlation renders communication complexity trivial~\cite{BBL+06}.
It achieves a temperally `flat' universal and blind MBQC~\cite{BFK09}. 
If the PR box is replaced by ebit, 
then an exponential amount of ebit is required to realize teleportation 
without communication~\cite{Vai03}.
However, the allowed amount of classical communication could also be finite instead of being minimal.
It remains to see if there is a quantum universal resource
that is stronger than MAGIC, but weaker than PMAGIC.

\section{The $h$-family} 
\label{sec:h-family}

\subsection{Hamiltonian quantum simulation}

In this section, we study a primary model based on Hamiltonian evolution.
In particular, we rely on the theory of a universal set $\C S$ 
of Hamiltonian interaction terms~\cite{CMP18,KPB+20,KPB+21,ZA21},
just like a universal gate set.
First, a $H'$ simulate $H$, up to local encoding $\C E$, below energy $\Delta$ if 
\be \| H'_{\leq \Delta} - \C E(H)\| \leq \epsilon \ee 
for $H'_{\leq \Delta}$ as the restriction of $H'$ up to energy $\Delta$.
More precise definitions can be found~\cite{CMP18,KPB+20,KPB+21}.
The simulation cost is a function of the system size and interaction energy of $H$.
Given $\epsilon$, the time evolution $U=e^{iHt}$ is simulated with errors up to $\epsilon t$.

A Hamiltonian quantum simulation process is formed by a sequence of local evolution
$e^{i t_n j_n h_n}$, with local terms $h_n$, parameters for interaction strength $j_n$
and time $t_n$. 
It can be viewed as a Lie-Trotter sequence. 
Each local term can be drawn from a set $\C S$, 
and a simulated Hamiltonian is constructed by a real-weighted sum 
\be H=\sum_n j_n h_n \ee
for amplitudes $j_n\in \B R$ and each term $h_n \in \C S$. 

It has been proven recently there are universal Hamiltonian 
sets~\cite{CMP18,KPB+20,KPB+21,ZA21},
such as the 2-local the Heisenberg and XY exchange interactions.
Given a target $H$, the simulation scheme first 
maps it to a quantum phase estimation (QPE) circuit $U$~\cite{ZA21},
and then uses Feynman-Kitaev circuit-to-Hamiltonian map to 
obtain a Hamiltonian $H_{FK}$~\cite{KSV02},
which is then simulated by a $\C S$-Hamiltonian $H'$ 
relying on the perturbation gadgets~\cite{BH17}.
It does not directly simulate $H$ in order to ensure the 
efficiency of the simulation, e.g.,
interaction amplitudes only grows polynomial with the system size. 

A special class of Hamiltonian is known to be stoquastic,
which has all off-diagonal elements being real and non-positive.
It is established that stoquastic Hamiltonian 
is more powerful than classical computation~\cite{CM16,BH17}.
This motivates our formulation of a QRT for HQS.
The free set $\C F$ is the set of stoquastic Hamiltonian, denoted as STOQ,
and free operations $\C O$ are any 
linear combination that preserves the stoquastic-ness, denoted as LINEAR.
For instance, the two-qubit stoquastic interaction is of the form 
$\alpha ZZ+ A\I + \I B$
for $\alpha\in \mathbb{R}$ in a proper basis~\cite{CM16,BH17},
while classical interaction is diagonal~\cite{lCC16}. 
A measure of the non-stoquastic-ness can be defined,
e.g., by distance or entropy based measures,
or by converting to the effects on states.
However, we would not explore this in this work,
and only refer to the particular form of the universal resources.




It is also valuable to draw the connection with and difference from other studies.
The subject of quantum simulation of Hamiltonian evolution aims to  
simulate a given evolution $U=e^{iHt}$ (or time-dependent ones) in the QCM
by decomposing it into a sequence of local unitary terms,
e.g., using Lie-Trotter formula, 
but without referring to a universal Hamiltonian set~\cite{Llo96,BACS07a}. 
HQS can be viewed as an extension of 
analog (dedicated special-purpose) quantum simulation~\cite{CZ12},
which has a weaker requirement on the controllability of local terms.
For instance, an analog simulator may have special form of local interaction terms,
the local switching on and off of which may not be available.
On the contrary, a HQS evolution is of digital natural since the set of $h_n$ is finite
and the time parameter can be digitalized, 
i.e., broken into controllable segments.
The requirement on time-dependence and classical control 
also makes it different from some Hamiltonian-based autonomous 
QC models~\cite{Jan07,Nag10,Nag12,CGW13,BHS+15,LT16,TGL+16},
such as continuous-time quantum walk~\cite{CGW13}. 
This reveals that the primary feature of the $h$ family 
is to explore the interactions among subsystems as resources.
Whether other requirements such as automation can lead to other family of models
is left for further investigation.

\subsection{Hamiltonian quantum cellular automata}

\begin{figure}
    \centering
    \includegraphics[width=0.1\textwidth]{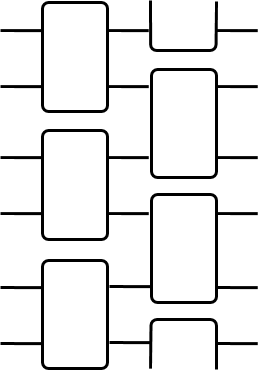}
    \caption{The schematics for a QCA brickwork circuit.}
    \label{fig:qca}
\end{figure}

When the arithmetic on Hamiltonian terms is restricted,
new models arise.
In this section, we define a HQCA model
which is a Hamiltonian-based QCA. 

For classical computatoin, CA is a universal model. 
It is known that there are 2D universal CA, 
while 1D CA cannot be universal~\cite{TM87,AB09}.
It arranges bits on a lattice with well-defined neighborhood, 
and evolution is specified by parallel local dynamics; e.g.,
the value of each bit at a later time is determined by its neighbors at present.
CA can be simulated by classical circuit. 
The local rule is mapped to a permutation $P$.
But before it, we need a COPY step for each bit.
The simulation circuit is of brickwork form, see Figure~\ref{fig:qca}, 
with a layer of COPY and a layer of $P$, and so on.

Diverse models of QCA have been developed~\cite{Arr19,Far20,Wie08}.
There are circuit-based and Hamiltonian-based ones, 
and here we use the later. 
Different from the classical case, 1D HQCA can be universal,
with two-local qudit intereaction terms. 
We first sketch a seminal model to show how it works,
and then introduce our model as a variation of it.
In the Nagaj-Wocjan model~\cite{NW08}, 
a local site with $d=10$ contains a few bands: for data qubit, program, and controller. 
The 1D chain is divided into many regions, with each region for a step in the simulated circuit. 
All the data qubits are located in a particular region.
A translational invariant interaction is designed to simulate the circuit.
Given a circuit of $N$ qubits and $M$ steps,
the HQCA system size is $MN$ and runs for a polynomial time.
Intuitively, the model works like a typing machine: 
data qubits do not move, while programs are executed by shifting the states of
the programs passing the data.
The model is autonomous, and as such
the desired final state $|\psi_f\ket$ is only 
realized with finite probability under the evolution $e^{iHt}$.
The success probability can be boosted by repeating the algorithm or
by modifying the model itself.



\begin{figure}[b!]
    \centering
    \includegraphics[width=0.15\textwidth]{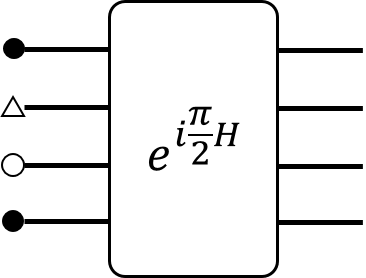}
    \caption{A local term in the HQCA model. The back dots are for data qubits,
    circle for an ancilla qubit, triangle for a program qutrit.}
    \label{fig:hqca}
\end{figure}

As for the HQS model, we allow classical control.
Here we define a classically controlled HQCA, borrowing idea 
from the classically controlled QCA~\cite{SFW06},
which often composes a sequence of different QCAs,
with each one controlled classically,
i.e., switched on and off. 
Our model is defined as follows.
For a 1D array of data qubits, 
add one ancilla qubit and program qutrit between each pair of data qubits.
The bold local dimension is twelve. 
The local interaction is 
\be H=|0\ket \bra 1| \otimes U + |1\ket \bra 0| \otimes U^\dagger, \label{eq:hqca}\ee
with the first part as the ancilla qubit,
and \be U=P_0 \otimes \I+P_1 \otimes W+P_2 \otimes \Pi, \label{eq:uqca} \ee
with the first part as the program qutrit,
for \be W=P_0 \otimes \I+P_1 \otimes HZ, \ee 
with the first part as a data qubit,
$\Pi$ as the SWAP gate on two data qubits. 
The gate $W$ is known to be universal 
if it can be applied to any pair of qubits~\cite{SFW06}.
Note in $HZ$ it is a Hadamard gate H.

Now given a circuit, composed with nearest-neighbor $W$ and SWAP gates,
first arrange it into a sequence of transversal steps.
Each step then is a tensor product of $W$ and SWAP gates.
The program qutrit $|p\ket$ encodes the type of gate, 
$G\in \{\I, W, \Pi\}$, acting on two data qubits.
The ancilla qubit is initialized to $|1\ket$.
A local evolution acts as
\be e^{i \frac{\pi}{2} H} |1\ket |p\ket |\psi\ket= |0\ket |p\ket G |\psi\ket,\ee 
which does not alter the program qutrit but flips the ancilla qubit.
The program qutrit and ancilla qubit needs to be reset before the next step.
The time $\frac{\pi}{2}$ is a constant so can be viewed as a fast quench,
and the whole HQCA is treated as Trotterized steps forming a brickwork structure.

Observe that if the $U$~(\ref{eq:uqca}) above is used instead of $H$~(\ref{eq:hqca}), 
then the ancilla qubit is not needed.
This basically reduces to a classically controlled QCA model~\cite{SFW06}.
If the program qutrit values are dragged out,
this will further reduce to the original circuit itself.
On the contrary, the main feature of the HQCA model is the parallelism.
One only needs to ensure the translational invariance of interaction.
Also, compared with the autonomous ones~\cite{NW08,VC08},
the use of classical control can ensure the deterministic preparation
of the desired final state. 

For a QRT of the HQCA model, it is natural to consider CA as the free set $\C F$.
The free operation $\C O$ is parallel control of local interactions, 
denoted as PARALLEL.
However, the gate $W$ does not reduce to the classical case apparently.
With the similar idea, we can pick the gate set of Toffoli and Hadamard gates,
and consider controlled-Toffoli and controlled-Hadamard gates as transversal steps,
although this will enlarge the locality of interaction and local dimension. 
As Toffoli is universal classically,
it is clear that the quantum resource is due to Hadamard gate,
i.e., it is coherence. 
Therefore, we identify COH as the universal resource for this model.
The relation between COH and NSTOQ can be seen by observing that
the local four-body term $H$~(\ref{eq:hqca}) is obviously non-stoquastic.
The term $H$ can be simulated in HQS by a weighted sum of two-body terms from a universal 
Hamiltonian set.

\subsection{Adiabatic quantum computing}

In this section we describe the AQC model~\cite{AL18}.
As it has been well known, 
we will merely show how it belongs to the $h$-family. 
In this model, usually one starts from the ground state $|\psi_0\ket$ 
of an easy-to-prepare $H_0$ as the initial state,
and then use adiabatic path 
\be H(t)=t H_0 + (1-t) H_1, \; t\in [0,1] \ee 
for the time-parameter $t$ and the ground state $|\psi_1\ket$ 
of $H_1$ encodes the final output.
The adiabatic condition requires the absence of gap-closing during the evolution 
$e^{i \int_0^1 H(\tau) d\tau}$,
which drives $|\psi_0\ket$ through the ground manifold $|\psi_t\ket$.
We can also attach a sequence of adiabatic paths together, in general.

The standard method to prove the universality of AQC 
is the Feynman-Kitaev history-state method.
Given a circuit $U=U_1 U_2\cdots U_L$, 
it is mapped to a five-local, but not geometrically, Hamiltonian $H_{FK}$
whose ground state is the history state
\be |\Phi\ket= \frac{1}{\sqrt{L+1}}\sum_{\ell=0}^L |\gamma_\ell\ket \ee 
for $|\gamma_\ell\ket=|\psi_\ell\ket|\ell\ket$, and
$|\psi_\ell\ket=U_\ell'|\psi_0\ket$, $U_\ell'=U_1U_2\cdots U_\ell$,
$|\psi_0\ket$ as the initial data state,
and $|\ell\ket$ as the state of a clock register.
An adiabatic path $H(t)$ is then design with $|\gamma_0\ket$ as the 
ground state of $H(0)$, and $|\Phi\ket$ as the 
ground state of $H(1)$. 
The history state only yields the output with probability $1/(L+1)$.
This is amplified by padding the circuit with a sequence of identity gates,
using more clock qubits and interaction terms, 
so that the success probability gets close to 1.




We now define the QRT for AQC belonging to the $h$-family.
We use on-site terms as free set $\C F$, 
corresponding to free product states, PRO,
which are often set as the initial state $|\psi_0\ket$ of an algorithm.
The free operation $\C O$, denoted as GAPPED,
is the adiabatic turning on and off of on-site terms,  
which is required not to induce gap-closing for each on-site term.
Indeed the adiabatic on-site switching 
is a special type of parallel or transversal switching. 
Then the universal resource is the terms that can lead to universal QC.
For the $H_{FK}$ model, 
restricted to the history manifold span$\{|\gamma_\ell\ket=|\psi_\ell\ket|\ell\ket\}$, 
the final Hamiltonian takes the form
\be H_w= \begin{pmatrix}
    \frac{1}{2} & -\frac{1}{2} &  &  \\
    -\frac{1}{2} & 1 & -\frac{1}{2} & \\
    & -\frac{1}{2} & 1 & -\frac{1}{2} & \\
    &  & \ddots & \ddots & \ddots  \\
\end{pmatrix}, \ee 
which is a 1D quantum walk model. 
Note that it is stoqustic but only in this special history-state basis.
A stoqustic Hamiltonian in the standard computational basis cannot be universal.
Therefore, we denote 1DQW as the universal resource for AQC.
It is obvious that, given more restrictive controllability,
the required interaction becomes nonlocal.
This is an echo of the resource theory for coherence and correlation
that we have established.
For resource conversion, it is clear 
the walk $H_w$ can be realized or simulated by QCA.
The basis transformation from $\{|\ell\ket\}$ to $\{|\gamma_\ell\ket\}$
is also simulated by QCA.






\section{Quantum algorithms and resources} \label{sec:alg}

In this section, we turn to present a primary 
resource-theoretic study of some well known quantum algorithms. 
We mainly focus on algorithms that rely on quantum circuits.
In particular, we study the interplay between coherence,
interference, entanglement, and contextuality.

\begin{figure}[b!]
    \centering
    \includegraphics[width=0.15\textwidth]{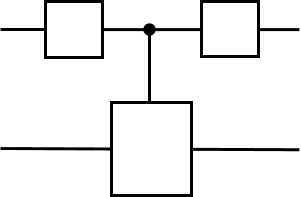}
    \caption{The schematics of a sandwiched circuit which is a special case
    of that in Fig.~\ref{fig:concir}.}
    \label{fig:sancir}
\end{figure}

For a quantum circuit, we can calculate the amount of interference.
We will use $I$ as the notation for interference. 
We use the relative entropy of coherence. 
In general, there is no definite relation between $I(V)$, $I(W)$,
and $I(VW)$ for two qudit gates $V, W \in U(d)$.
If $V$ is incoherent, then $I(VW)=I(WV)=I(W)$.
If an incoherent $V$ is also nonunitary,
then $I(VW) \leq I(W)$, $I(WV) \leq I(W)$, and 
$I(V_2WV_1) \leq I(W)$ for incoherent $V_1$ and $V_2$.
An important circuit form is the sandwiched one,
shown in Figure~\ref{fig:sancir}, with the
controlled-$U$ gate as a multiplexer defined by Eq.~(\ref{eq:plex}).
The control and data registers can be of different dimensions,
$d_1$ and $d_2$.
The interference of $CU$ is the average $\frac{1}{d_1}\sum_i I(U_i)$.
For $(V\otimes \I)CU(W\otimes \I)$, the relative entropy is the average
\be \frac{1}{d_1d_2} \sum_{b,\nu} H(p_{a,\mu}) \ee 
for $p_{a,\mu}=|\sum_i v_{ai}w_{ib} U_{i,\mu\nu}|^2$.
The sum over $i$ is an interference.
A simple special case is when $CU$ is classical,
then it reduces to $I(VW)$.
Also we find the additive relation
\be I(CU(V\otimes \I))=I((V\otimes \I)CU)=I(V) + I(CU).\ee 
For instance,
$I(CX)=0$, $I(CX(U\otimes \I))=I(U)$, $I(CU(H\otimes \I))=1+I(CU).$
This confirms our intuition.
Many algorithms have the sandwiched form of circuit, 
such as the DQC1 and SWAP-test~\cite{KL98,BCW+01}.
Our study shows that interference is the source for their computational power.

\subsubsection{Van den Nest's model}

In the circuit model, an algorithm on $n$ qubits starts from $|0\ket^{\otimes n}$ 
and evolves a circuit $U$,
and the output is obtained by measuring the first qubit in the computational basis
with $p_0$ as the success probability.
Given this, Van den Nest's model converts to a circuit with one additional qubit,
and the final state before measurement takes the form
\be |\psi\ket= \sqrt{1-\epsilon} |0\ket|0\ket^{\otimes n} 
+\sqrt{\epsilon}|1\ket U|0\ket^{\otimes n}, \ee
for $\epsilon\sim 1/\text{poly}(n)$ as a small error parameter
so that the success probability can be boosted efficiently.
It was shown that the entanglement is polynomially small,
while this model is universal. 

We will show that the amount of interference in this model is not small.
Let $V_\epsilon=
\begin{pmatrix} \sqrt{1-\epsilon} & -\sqrt{\epsilon} \\ 
\sqrt{\epsilon} & \sqrt{1-\epsilon} \end{pmatrix}$.
The interference of the Van den Nest circuit is
\be I(CU(V_\epsilon\otimes \I))=I(V_\epsilon)+\frac{1}{2} I(U) \ee
for $I(V_\epsilon)$ as $C_r(V_\epsilon)=h(\epsilon)$ or $C(V_\epsilon)=2\sqrt{\epsilon(1-\epsilon)}$,
which are close to zero.
The major contribution is from the circuit itself $I(U)$,
which means the feature of the circuit for solving a problem is preserved.

Therefore, we see by the map from $U$ to Van den Nest's circuit,
the entanglement may change significantly,
while the interference does not.
This explains why this model works.
However, it was argued that entanglement is neither necessary nor sufficient for 
quantum speedup. 
Our resource-theoretic study denies this, 
and it shows that a universal resource needs to be defined in a UQCM.
We showed that entanglement or EBIT is the universal resource for LQTM,
which is closely related to tensor-network states.
As analyzed in Section~\ref{subsec:lqtm},
entanglement mainly refers to the quantum correlation between parts,
while coherence and interference refers to the dynamical change of the amplitudes.
In other words, we can put entanglement and interference as orthogonal axis in a coordinate,
with entanglement describing static feature of a state, 
while interference describing how a state evolves.
However, entanglement and coherence are also closely related. 
A crucial fact is that in Van den Nest's model entanglement is only polynomially small,
which can be boosted efficiently, 
while an exponentially small one cannot be.
This means entanglement is indeed still there, 
and it is more convenient to use interference instead to describe its feature.





\subsubsection{Linear combination of unitary algorithm}

In recent years, the LCU algorithm has been developed~\cite{CW12,BCC+15,WL16,Long06,Long11}.
Originally, LCU was motivated by the 
multi-slit interference experiment:
a particle, which has two degree of freedoms,
namely, path $\C H_p$ and spin $\C H_s$, will follow the interference pattern observed on a special screen.
The interference is for the particle as a whole.
In a LCU task 
\be U=\sum_i c_i U_i, \ee 
the $\sum_i c_i $ is from the path,
but $U_i$ acts on the spin. 
Its primary component also has the sandwiched form of circuit,
probably followed by post-selection.

A caveat is that the interference here is not for amplitudes of states,
instead, it is for unitary operators.
Actually, we have argued in Section~\ref{subsec:cont} that this relates to contextuality.
The form $\sum_i c_i U_i$ is a quantum superposition of contexts 
each defined by a $U_i$.
This could be a better point of view for LCU
and algorithms with the sandwiched circuit.
Indeed, the sandwiched circuit is not the one with the maximal amount of interference
for the whole space $\C H_p \otimes \C H_s$.
A Hamiltonian evolution $e^{itH}$ or Fourier transformation 
on the whole space can have larger amount of interference,
which can also show quantum speedup,
e.g., the quantum walk on special graphs~\cite{CCD+03}. 
Therefore, we see that although contextuality and coherence are equivalent universal resources,
their usages in algorithms could be different. 
This provides the flexibility for the design of algorithms in different settings.







\subsubsection{Shor, Grover, \emph{et al.} algorithms}

Shor's factorizing algorithm has been one of 
the most significant progresses 
for quantum computing~\cite{Sho94}. 
It is closely related to the algorithms by Deustch-Jozsa, 
Simons, and the quantum phase estimation by Kitaev,
and solves problems in the class of hidden subgroup~\cite{NC00}. 
Besides some classical side-processing,
the primary quantum circuit is of the sandwiched form,
which is a contextual circuit from Def.~\ref{def:contcir}.
Therefore, the resource can be attributed as interference or contextuality. 
In fact, contextuality is more apparent than interference: 
the interference of the control register, which enables contextuality, is more crucial. 
The black-box unitary in the modular exponentiation behaves classically.
The quantum Fourier transform is used on the control register
whose interference is maximal.

The amount of interference in Deustch-Jozsa algorithm is not extensive,
so it does not have exponential speedup over random algorithms.
For Simons and Shor algorithm, the final state is in superposition,
and the interference is extensive. 
There is also an extensive amount of entanglement. 

From these algorithms, it is clear to see how interference works
as the arithmetic of the amplitude.
The ``amplitude arithmetic'' is the additional quantum part beyond classical algorithms.
In particular, amplitude can be negative, hence can lead to destructive interference.
A quantum speedup occurs when a success probability is boosted by interference based on the amplitude arithmetic.




Grover's search algorithm~\cite{Gro96} 
demonstrated the power of a different type of quantum algorithms.
Despite its oracle setting,
it is better described as a qubit rotation in a suitable basis.
The rotation angle increase linearly with its iteration. 
It appears that the amount of interference is not extensive
for a qubit evolution~\cite{BG06,Sta14}. 
However, we have to take into account of the basis which 
would mostly not be the computational basis. 
As coherence is defined relative to the computational basis, 
there could be a large amount of interference in Grover's algorithm. 

Grover's algorithm can be viewed as a state-preparation algorithm.
For instance, for a problem to prepare $|\psi\ket$ which satisfies 
$|\psi\ket=U|0\ket$, the interference of $U$ is characterized by the coherence of $|\psi\ket$.
For generic state $|\psi\ket$, 
the quantum speed limit from the uncertainty principle
can be used to lower bound the time needed for its preparation~\cite{LT09}.
This also provides the lower bound $\Omega(\sqrt{N})$ for the search problem 
of an unstructured database of size $N$, and proves 
the optimality of Grover's algorithm~\cite{FG96}.

Recently, 
Grover's algorithm has been generalized via ``qubitization'' and 
quantum singular-value transformation (QSVT)~\cite{GSLW19,MRTC21}. 
Without going into the details,
a unitary $U$ is treated as a direct sum of qubit-rotations $U_i$ 
each for a singular value $s_i$ in a special basis.
Grover's algorithm is the case with only one singluar value.
The interference of $U$ is therefore the average $\sum_i I(U_i)/d$,
hence is not extensive.
But the coherence for this special basis needs to be considered.
This leads to the potential of an extensive amount of interference 
and a quantum speedup by the QSVT algorithm. 



\section{Conclusion} \label{sec:conc}

In this work, we studied families of universal quantum computing models (UQCMs)
using quantum resource theory (QRT).
We have shown that QRT offers a rigorous framework to
characterize a UQCM and even classify them,
and UQCMs serve as broad settings to utilize resources
and explore quantum primacy. 
For each family, we identified a triplet of models.
This is not unique and is likely the smallest set of generations. 



The quantum circuit model (QCM) indeed is simple and easy to use.
It lies at the lowest level in the amplitude family.
However, QCM is often found not enough to satisfy more realistic or theoretical needs,
such as security, programmability, modularity, controllability, energy efficiency, 
and parallelism, etc. 
This requests a diverse exploration of quantum resources and UQCMs.
For further investigation, we would like to remark on a few points. 

We find many models can be identified by a particular form of circuit.
In this work, we have analyzed the transversal,
sequential, brickwork, sandwiched, contextual, tailed, and stabilizer circuits.
Theses circuits are motivated by the settings of QRT,
and each may be suitable to satisfy a particular needs.
They can also be used for other purposes,
such as quantum speedup~\cite{BGK18} and the classification
of entangled states. 

Among the nine UQCMs that we studied, 
the contextual quantum computing (CQC) 
and post-magical quantum computing (PMQC)
are relatively new. 
The CQC model relies on our notion of contextuality.
However, the landscape of contextuality is a maze~\cite{BCG+21}.
There are also schemes or models that utilize contextuality,
so it remains to see how these contextual schemes 
relate and if they can be unified for a consistent notion of contextuality.
The PMQC model is the only one that is slightly beyond quantum theory
due to the nonlocal box~\cite{PR94}.
It is not clear if there is a weaker model than PMQC on one hand,
and on the other hand, 
the understanding of the nonlocal world itself~\cite{DGH+07} is also incomplete.

Our study of the Hamiltonian family is relatively less in depth. 
We did not study measures of resources and algorithms.
However, it implies that 
this family is equally powerful with other families,
and it is worthy to study Hamiltonian-based resources, 
schemes for fault-tolerance, etc. 
Also this family does not request automation,  
i.e., a free evolution $e^{iHt}$ without control in the middle.
Such autonomous Hamiltonian-based models are also shown to be 
powerful~\cite{Jan07,Nag10,Nag12,CGW13,BHS+15,LT16,TGL+16},
hence can be studied further. 

We mentioned that, besides the three families we studied,
there should also be other families,
such as the evolution family, a coding-based family,
and even non-universal family for restricted purposes
such as communication, metrology, etc~\cite{KCS+20}. 
In particular, the evolution family relies on the set of 
quantum channels, the arithmetic on which are known as 
combs or superchannels~\cite{CAP08}.
The recently proposed quantum von Neumann architecture~\cite{Wang22} 
relies on this and can execute quantum superalgorithms,
which can automate the design of a quantum algorithm by another one. 
We will leave this for future work.\\ \vspace{-1cm}

\section*{Acknowledgements}

This work has been funded by
the National Natural Science Foundation of China under Grants number
12047503 and number 12105343.

\end{spacing}

\bibliography{ext}{}
\bibliographystyle{ieeetr}


\end{document}